\documentclass[prb,twocolumn] {revtex4-1}

\usepackage[english]{babel}

\makeatletter
\def\bbl@set@language#1{%
	\edef\languagename{%
		\ifnum\escapechar=\expandafter`\string#1\@empty
		\else\string#1\@empty\fi}%
	\@ifundefined{babel@language@alias@\languagename}{}{%
		\edef\languagename{\@nameuse{babel@language@alias@\languagename}}%
	}%
	\select@language{\languagename}%
	\expandafter\ifx\csname date\languagename\endcsname\relax\else
	\if@filesw
	\protected@write\@auxout{}{\string\select@language{\languagename}}%
	\bbl@for\bbl@tempa\BabelContentsFiles{%
		\addtocontents{\bbl@tempa}{\xstring\select@language{\languagename}}}%
	\bbl@usehooks{write}{}%
	\fi
	\fi}
\newcommand{\DeclareLanguageAlias}[2]{%
	\global\@namedef{babel@language@alias@#1}{#2}%
}
\makeatother

\DeclareLanguageAlias{en}{english}

\usepackage{amsmath,amssymb,graphicx}

\usepackage[colorlinks=true,linkcolor=blue,urlcolor=blue,citecolor=blue,breaklinks=true]{hyperref}
\usepackage[utf8]{inputenc}

\usepackage{color}

\newcount\minute
\newcount\hour
\newcount\hourMins
\def\now
{
  \minute=\time
  \hour=\time \divide \hour by 60
  \hourMins=\hour \multiply\hourMins by 60
  \advance\minute by -\hourMins
  \zeroPadTwo{\the\hour}:\zeroPadTwo{\the\minute}
}\def\timestamp
{
  \today\ \now
}
\def\today
{
  \zeroPadTwo{\the\day}-\zeroPadTwo{\the\month}-\the\year%
}
\def\zeroPadTwo#1%
{% Left zero pad of the argument to 2 digits.  The argument
  \ifnum #1<10 0\fi
  #1%
}
\date{\timestamp -\jobname.tex}

\begin{document}

\title{Microscopic theory of the friction force exerted on a quantum impurity in one-dimensional quantum liquids}

\author{Aleksandra Petkovi\'{c}}
\affiliation{Laboratoire de Physique Th\'eorique, Universit\'e de Toulouse, CNRS, UPS, France}

\begin{abstract}
We study the motion of a slow quantum impurity in one-dimensional environments focusing on systems of strongly interacting bosons and weakly interacting fermions. While at zero temperature the impurity motion is frictionless, at low temperatures finite friction appears. The dominant process is the scattering of the impurity off two fermionic quasiparticles. We evaluate the friction force and show that, at low temperatures, it scales either as the fourth or the sixth power of temperature, depending on the system parameters. This is a result of the scattering of the impurity off two fermionic quasiparticles that are situated around different Fermi points. It is the dominant process at low temperatures. We also evaluate the contribution to the friction force originating from the scattering of the impurity off two fermionic quasiparticles that are situated around different Fermi points. It behaves as the tenth power of temperature.
\end{abstract}


\maketitle
\section{Introduction}

Understanding the motion of a quantum particle through a medium is a fundamental problem relevant for a large class of systems \cite{landau+49,Devreese_2009,SCHECTER2012639}. Substantial experimental progress in controlling and fabricating distinguishable particles (impurites) in quantum liquids offers new perspectives and provides an ideal playground for this problem \cite{chikkatur2000suppression,palzer2009quantum,PhysRevLett.102.230402,
PhysRevLett.103.170402,zipkes2010trapped,schmid2010dynamics,
weitenberg2011single-spin,catani2012quantum,
PhysRevLett.109.235301,NaturePolarons,NaturePolaronsII,
fukuhara2013quantum,PhysRevLett.117.055302,
PhysRevLett.117.055301,PhysRevLett.118.083602,Cetina96,NatureCasimir}. One-dimensional liquids are a particularly interesting environment due to  pronounced effects of quantum fluctuations and correlations \cite{mcguire1965interacting,yang1967exact,gaudin1967systeme, castella1993exact,castro_neto1996dynamics,
fuchs2005spin,zvonarev2007spin,matveev2008spectral,
lamacraft2009dispersion,
kamenev2009dynamics,gangardt2009bloch,
matveev2012scattering,
bonart2012nonequilibrium,peotta2013quantum,burovski2014momentum,
knap2014quantum,PRLimpurity,robinson2016motion,PRBBose-Fermi,Meinert945,PhysRevA.96.043625,PhysRevLett.121.080405,Casimir,PRBCasimir,Pasek,Wall+impurity,PhysRevLett.122.183001}.

At zero temperature, a slow mobile impurity in a one-dimensional liquid experiences frictionless motion \cite{landau+49,castro_neto1996dynamics}. At low temperatures, a finite friction force appears  due to impurity scattering off thermally excited quasiparticles of the liquid. An early study of  \citet{castro_neto1996dynamics} showed that the resulting  friction force scales as the fourth power of  temperature. This scaling is valid both for bosonic and fermionic liquids. Later,  using a phenomenological approach the prefactor of this $T^4$ law was expressed in terms of the  chemical potential of the liquid, the chemical potential of the impurity, and effective impurity mass and their derivatives with respect to the liquid density \cite{gangardt2009bloch,matveev2012scattering}. However, the determination of the chemical potentials and the effective impurity mass is a difficult task in a general case.

Recently, in Ref.~\cite{PRLimpurity} the friction force was examined in detail for the system of weakly interacting one-dimensional bosons in a wide range of temperatures. At low temperatures, it was found that by tuning the system parameters the coefficient in front of the $T^4$ law can vanish resulting in $T^8$ dependence of the friction force. For temperatures above the chemical potential of the Bose gas, the linear  dependence on temperature was found. The analytic expression for the friction force was obtained in the crossover region between the low-temperature and high-temperature regimes \cite{PRLimpurity}.

Usually low-energy properties of one-dimensional systems are studied within the Luttinger liquid theory that assumes linear dispersion relation of quasiparticles. However, in order to determine functional dependence of the friction force on system  parameters one has to go beyond this description even at lowest temperatures \cite{gangardt2009bloch,matveev2012scattering,
PRLimpurity}. Since complete analysis of the nonlinear spectrum is a very demanding task, in this paper we focus on systems of strongly interacting bosons and weakly interacting fermions. This allows us to develop a microscopic theory of the friction force  and derive new laws. Our results are as follows. In Sec.~\ref{model} we introduce the model and calculate the scattering matrix element for slow impurity immersed in a system of weakly interacting fermions. We consider an arbitrary form of the two-body interaction. In Sec.~{\ref{sec:General}} we evaluate the friction force assuming that the Fourier transform of the interaction is given by an analytic function. We are interested in  temperatures smaller than the Fermi energy.  We demonstrate that by controlling the system parameters one can design the desired friction in the system. Namely, we show that the friction force can dramatically change its temperature dependence from the expected $T^4$ behavior \cite{castro_neto1996dynamics,gangardt2009bloch,matveev2012scattering} to a new $T^6$ law. This is the result of the scattering of the impurity off two fermions that are situated around different Fermi points, while the scattering of the impurity off fermions that are around the same Fermi point leads to the contribution to the friction force that is proportional to $T^{10}$. 
In Sec.~\ref{noninteracting} we apply the obtained results to the impurity immersed in a gas of Tonks-Girardeau bosons or equivalently in a system of non-interacting fermions. The interaction between the impurity and the background particles is assumed to be the contact interaction. When the mass of the impurity equals the mass of the background particles, the system is integrable \cite{mcguire1965interacting,castella1993exact} and the impurity becomes transparent for the background particles, leading to the absence of the friction force \cite{Zotos2002}. We find that only for an impurity much lighter than the background particles, the $T^{10}$ contribution to the force may become the dominant one at higher temperatures that remain below the Fermi energy. In Sec.~\ref{strong} we consider a strongly interacting boson environment  that can be mapped \cite{PhysRevLett.82.2536,PhysRevLett.99.110405} onto a weakly interacting Cheon-Shigehara model of fermions. Section \ref{screened} focuses on fermions interacting via the screened Coulomb interaction. This system does not satisfy the assumptions of Sec.~\ref{sec:General} and requires separate analysis. 
The discussion and conclusions are given in Sec.~\ref{conclusions}, while some technical details are presented in appendices.

%%%%%%%%%%%%
\section{Model}\label{model}
%%%%%%%%%%%%

We consider the motion of a mobile impurity immersed in a one-dimensional system of spinless fermions. The model is given by  the Hamiltonian
\begin{align}\label{Ham}
H=H_0+H'.
\end{align}
Here $H_0$ contains the kinetic energy of fermions of mass $m$ and of a single impurity of mass $M$ that in second quantized  form read as\begin{align}
H_0&=\sum_p \frac{p^2}{2m}a_p^\dag a_p+ \sum_P\frac{P^2}{2M}B^{\dag}_PB_P.
\end{align}
The impurity and the fermion creation and annihilation operators are denoted by $B^\dag$ and $B$ and by $a^\dag$ and $a$, respectively. Since we consider a single impurity, one has $\int dx B^{\dag}(x)B(x)=1$.
The fermions interact between themselves and with the impurity. The interactions take the form
\begin{align}
H'&=\frac{1}{L}\sum_{P,p,q}G_{q}B^{\dag}_{P+q}B_{P}a^{\dag}_{p}a_{p+q}\notag\\&+\frac{1}{2L}\sum_{p_1,p_2,q}V_q a^{\dag}_{p_1+q}a^{\dag}_{p_2-q}a_{p_2}a_{p_1}.
\end{align}
Here $V_q$ and $G_q$ denote the Fourier transforms of the two-body interaction between fermions and between fermions and the impurity, respectively. The fermionic operators satisfy standard anticommutation relations: $\{a_p,a^\dag_q\}=\delta_{p,q}$ and $\{a^{\dag}_p,a^{\dag}_q\}=\{a_p,a_q\}=0$. In what follows, we consider a weakly interacting system at temperatures well below the Fermi energy. In this case the scattering processes can be classified by the number of fermions involved. The condition on interactions is $G_0\ll \hbar v_F$ and $\left | V_0-\int_{-1}^{1}dx\int_{-1}^{1}dy V_{|x-y|m v_F}/4\right |\ll \hbar v_F$,  where $v_F$ denotes the Fermi velocity. We will study different types of the interaction potential. We point out that we discuss finite mass impurities that are very different from the infinitely heavy impurities considered in Refs.\cite{PhysRevB.46.15233,PhysRevE.55.2835,PhysRevA.66.013610,astrakharchik2004motion,Cherny2012,sykes}.

Our goal is to evaluate the friction force exerted on the impurity. Let us consider a process where the impurity of momentum $P$ scatters off a single fermion with momentum $p$ and acquires the momentum $P+\delta P$ while the final fermion momentum is $q$. The interactions will be taken into account within the perturbation theory. From the energy and the momentum conservation of a free system it follows that $
p={mP}/{M}+{\delta P}/{2}\left({m}/{M}+1\right)$ and
$q={mP}/{M}+{\delta P}/{2}\left({m}/{M}-1\right)$, 
where $p_F$ denotes the Fermi momentum. At zero temperature, the Pauli principle imposes the constraints: $|p|\leq p_F$ and $|q|\geq p_F$. One thus obtains that  the scattering process is not allowed \cite{PhysRevE.90.032132} at zero temperature provided the impurity momentum satisfies
\begin{align}\label{condition}
|P|/p_F<\mathrm{min}\{1,M/m\}.
\end{align} 
In what follows we assume that this condition is fulfilled. At finite temperatures, the above constraints on the momenta of the initial and the final fermions are replaced by the product of two Fermi distributions $n_{p}(1-n_{q})$, where $n_p=1/(1+e^{(\epsilon_p-E_F)/k_BT})$ and $E_F$ is the Fermi energy. However, by taking into account that the bare quadratic dispersion relation of the impurity gets strongly renormalized at momenta greater than or similar to  $p_F \mathrm{min}\{1,M/m\}$\cite{Schecter_2016}, the single-fermion process remains forbidden at finite temperatures and the dominant contribution to the friction force arises from the scattering off two fermions \cite{castro_neto1996dynamics}.

Next we analyze the processes involving two fermions. We calculate the scattering matrix element defined as a vacuum expectation value $t_{P,p1,p2}^{Q,q1,q2}=\langle B_Q a_{q_2}a_{q_1}|T|B_P^{\dag}a^{\dag}_{p_1}a^{\dag}_{p_2}\rangle$, where the $T$-matrix can be expanded as  $T=H'+H'\frac{1}{\langle i |H_0| i \rangle-H_0+ i 0^{+}} H'+\ldots$. Here the inital state is denoted by $|i\rangle =|B_P^{\dag}a^{\dag}_{p_1}a^{\dag}_{p_2}\rangle$. 
After introducing $t_{P,p_1,p_2}^{Q,q_1,q_2}=t\delta_{p_1+p_2+P,q_1+q_2+Q}$ we get
%\begin{widetext}
\begin{align}\label{t}
t=&[1-\hat{\mathcal{A}}(p_1,p_2)][1-\hat{\mathcal{A}}(q_1,q_2)]\frac{1}{L^2}\notag \\&\times\Bigg[\frac{G_{Q-P}V_{p_1-q_1}}{\epsilon_{p_1}+\epsilon_{p_2}-\epsilon_{q_1}-\epsilon_{q_2+Q-P}}\notag\\&+\frac{G_{Q-P}V_{p_2-q_2}}{E_P+\epsilon_{p_1}-E_Q-\epsilon_{p_1+P-Q}}\notag\\&+\frac{ G_{p_2-q_2}G_{p_1-q_1}}{E_P+\epsilon_{p_1}-E_{Q+q_2-p_2}-\epsilon_{q_1}}
\Bigg]
\end{align}
%\end{widetext}
in the leading order in the interactions. Here the operator $\hat{\mathcal{A}}$ is defined as $\hat{\mathcal{A}}(p_1,p_2)f(p_1,p_2,q_1,q_2)=f(p_2,p_1,q_1,q_2)$. We assumed that both $p_1$ and $p_2$ are modified in the scattering process. The energy dispersion of fermions is denoted by $\epsilon_{p}=p^2/2m$, while the impurity dispersion is $E_P=P^2/2M$. 

Notice that the special case of Eq.~(\ref{Ham}), i.e.,  the system of noninteracting fermions ($V_q=0$) coupled by a contact interaction with an impurity of the same mass as fermions (i.e.~$M=m$) is an integrable model  \cite{mcguire1965interacting,castella1993exact}. One thus expects the absence of the impurity scattering off the quasiparticles \cite{Zotos2002}. Indeed, the scattering matrix element (\ref{t}) vanishes once the energy and the momentum conservation laws are taken into account.  

The scattering matrix element (\ref{t}) determines the impurity dynamics. In the following we focus on the friction force exerted on the impurity by the liquid. It is given by Fermi's golden rule
\begin{align}\label{force}
F=\frac{2\pi}{\hbar}\sideset{}{'}\sum_{p_1,p_2,Q,q_1,q_2} \left|t_{P,p1,p2}^{Q,q1,q2}\right|^2 (Q-P)n_{p_1}n_{p_2}(1-n_{q_1})\notag\\(1-n_{q_2})\delta{(E_P+\epsilon_{p_1}+\epsilon_{p_2}-E_Q-\epsilon_{q_1}-\epsilon_{q_2})}. 
\end{align}
We point out that the symbol $\sum\nolimits'\ $ denotes that the sum is taken over distinct inital and final physical states. Next we analyze this expression. We consider that the impurity has a positive momentum. Additionally, we assume that the impurity is slow and satisfies $P\ll p_F$. 
The product of Fermi distributions in (\ref{force}) and the conservation laws impose that at low temperatures the fermions in the initial and in the  final state have to be in the vicinity of the Fermi points.  There are three types of configurations to be distinguished regarding the initial state. It can consist of i) one fermion with positive momentum and one with negative momentum; ii) two fermions with positive momenta and iii) two fermions with negative momenta. The configurations i), ii) and iii) are characterized by the total momentum close to zero, $2p_F$ and $-2 p_F$, respectively. Since the impurity is slow, the momentum conservation imposes that the  configuration of fermions in the final state has to be of the same type as the initial one. For example, if the incoming fermions have negative momenta the outgoing ones  also have negative momenta. Thus we have to distinguish the three above mentioned cases. We refer to them as the i) $1-1$, ii) $0-2$ and iii) $2-0$ scattering processes, see Fig.1. All of them give a nonzero contribution to the friction force only at finite temperatures. We will study them in the following sections.

\begin{figure}
\includegraphics[width=0.95\columnwidth,angle=0]{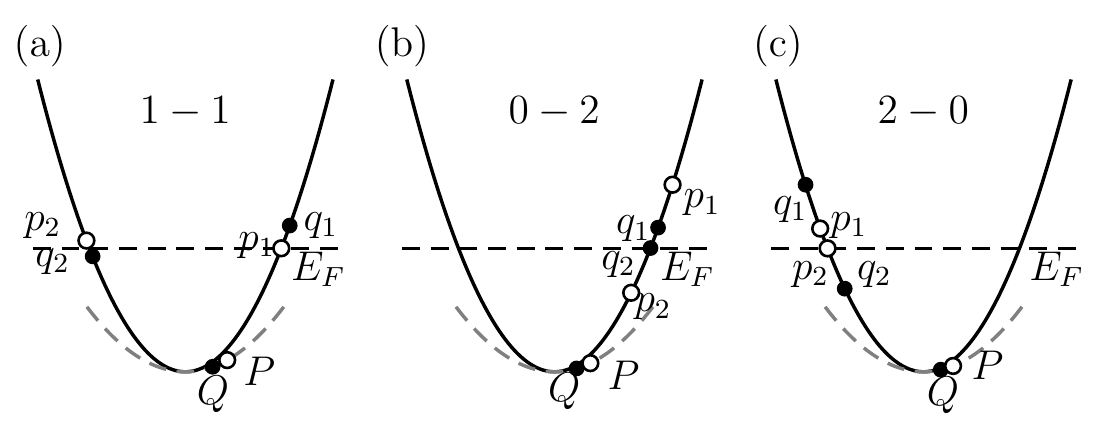}
\caption{The impurity of momentum $P$ scatters off two fermions with momenta $p_1$ and $p_2$. Schematic depiction of three different type of processes (a) $1-1$, (b) $0-2$, and (c) $2-0$ that lead to finite friction force at nonzero temperatures. Empty and filled circles represent, respectively, incoming and outgoing momenta. Solid and dashed parabolas show the energy of fermions and the impurity, respectively, as a function of momentum. }
\end{figure}

\section{Friction force}\label{sec:General}

In this section we evaluate the scattering matrix element (\ref{t}) and the friction force (\ref{force}) assuming that the Fourier transform $V_q$ of the interaction between the fermions is an analytic function of $q$. This allows us to express the friction force in terms of $V_q$ and its derivatives. We assume that the interaction between the impurity and the background particles takes form of the contact interaction, i.e., $G_q=G$. 

\subsection{$1-1$ processes}

We start by analyzing different scattering processes and their contributions to the friction force. We first consider the $1-1$ type processes. We assume that $p_1$ and $q_1$ are positive, while $p_2$ and $q_2$ are negative, see Fig.1a. There are other configurations of fermions of the $1-1$ type giving the same result. However, they do not provide distinct initial and final physical states of fermions and thus do not contribute in Eq.~(\ref{force}). We express the final momentum of the impurity as  $Q=P+\delta P$. At low temperatures the initial momenta are around the Fermi momentum and one has $|p_2+p_F|\sim|p_1-p_F|\sim T/v_F$. Analyzing the product of distribution functions in Eq.~(\ref{force}) and taking into account the energy and the momentum conservation we get that $|\delta P|\sim T/v_F$. Further,  the scattering matrix element (\ref{t}) can be simplified by taking advantage of the smallness of $T/E_F\ll1$ and  in the leading order in $T$ it is
\begin{align}
\label{t11General}
t^{1-1}&=G M\frac{G (M^2-m^2)+m M (V_{0}-V_{2 p_F}+2 p_F V'_{2 p_F})}{L^2 (M^2p_F^2-m^2P^2)}.
\end{align}
The energy conservation can be expressed as
\begin{align} \label{E1-1}
\delta{(E_i-E_f)}\approx \frac{1}{2v_F} \delta\left(q_1-p_1+\delta P\frac{V+v_F}{2 v_F}\right)
\end{align}
where $E_i=E_P+\epsilon_{p_1}+\epsilon_{p_2}$ and $E_f=E_Q+\epsilon_{q_1}+\epsilon_{q_2}$. Here we introduced  the impurity velocity $V=P/M$. 
The relation (\ref{E1-1}) and the momentum conservation allow us to express $q_1$ and $q_2$ in terms of the initial fermion momenta and $\delta P$. Thus it remains to evaluate the three summation over $p_1$, $p_2$ and $\delta P$ in Eq.~(\ref{force}). As discussed above, each summation is restricted to a momentum range of the width determined by $T/v_F$. Thus the phase space volume is proportional to $T^3$. Taking into account the typical momentum change $\delta P\sim T/v_F$ and the temperature-independent squared scattering matrix element $t^{1-1}$ in Eq.~(\ref{force}), we get that the friction force is proportional to $T^4$. Indeed, the detailed calculation of the force give us 
\begin{align}\label{TfourthGeneral}
F^{1-1}=&-\frac{2 \pi }{15\hbar ^5}\frac{G^4 T^4 V (M^2-m^2)^2 ( v_F^2+V^2) }{M^2m^4
   v_F^2  \left(v_F^2-V^2\right)^5}\notag\\ &\times\left(1 +\frac{m M \left(V_0-V_{2 m
   v_F}+2 m v_F V'_{2 m v_F}\right)}{G(M^2-m^2)} \right)^2
\end{align}
The temperature is assumed to be $T\ll E_F\text{min}\{1,M/m\}$. Note that in the case of non-interacting fermions when the impurity mass equals the mass of fermions, $M=m$, the system is integrable\cite{mcguire1965interacting,castella1993exact} and thus the friction force vanishes. Some useful integrals needed for calculation of the expression (\ref{TfourthGeneral}) are presented in App.~\ref{appendix}. 

We see that the perturbation theory breaks down when the impurity velocity approaches the Fermi velocity. This parameter region is indeed out of the domain of validity of our theory, since we study slow impurity. The reason is that at momenta greater than or similar to  $p_F \mathrm{min}\{1,M/m\}$ the renormalized impurity dispersion significantly deviates from the quadratic one\cite{Schecter_2016}, $E_P=P^2/2M$. Thus, in the expression (\ref{TfourthGeneral}), the impurity velocity $V$ should be replaced by the renormalized impurity velocity\cite{matveev2012scattering}  that is always smaller than $v_F$. As a result, the modified Eq.~(\ref{TfourthGeneral}) would be always finite.

Regarding the constraints on temperature, we notice that at temperatures greater than $E_F M/m$ thermal fluctuations become very strong and for this specific process  the typical change of the impurity momentum is big $\delta P \gtrsim P$. The fluctuations are such that the impurity can  be excited to momenta $\sim p_F M/m$ that can not be described by our theory.
For a heavy impurity the result (\ref{TfourthGeneral}) holds for $T\ll E_F$, however for a light impurity it applies for $T\ll E_F M/m$. 

We point out that the low-temperature friction force experienced by the slow impurity immersed in a system of weakly interacting bosons described by the Lieb-Liniger model has the same dependance on the impurity velocity as that in Eq.~(\ref{TfourthGeneral}) once the Fermi velocity is replaced by the sound velocity\cite{PRLimpurity}. We will show in the following sections that this property remains valid also for the Tonks-Girardeau gas, the Lieb-Liniger model with finite but strong interaction potential  as well as for fermions interacting via weak the screened Coulomb interaction. We expect this to be the universal property of one-dimensional systems. 

We emphasize that one can tune interactions such that contributions originating from the impurity--liquid coupling and the interactions between the fermions cancel each other in Eq.~(\ref{TfourthGeneral}). This is realized for  
\begin{align}\label{condition4}
G (M^2-m^2)+m M (V_{0}-V_{2 p_F}+2 p_F V'_{2 p_F})=0.
\end{align}
If the system is not integrable when the condition (\ref{condition4}) is satisfied, the scattering matrix element is expected to be nonzero. Thus, we should expand Eq.~(\ref{t}) to the next order in temperature. We obtain the following expressions
%\begin{widetext}
\begin{align}\label{t1-1spec}
t^{1-1}={}&\frac{G}{2 m L^2 p_F^4 (m^2 P^2 - M^2 p_F^2)}\notag\\ &\times\left[\alpha (p_1-p_F)+\beta (p_2+p_F)+\gamma \delta P \right],\\
{\alpha}={}& [G P(M^2-m^2)  (m P + M p_F) - 
   2 m^2 M p_F^4 V''_{2p_F}]\notag\\ &\times 2 M p_F,\\
{\beta}={}&- [GP (M^2-m^2)  (m P - M p_F) - 
   2 m^2 Mp_F^4  V''_{2p_F}]\notag\\ &\times {2 M p_F},\\
\gamma={}& G (m^2 - M^2) P [m^2 P^2 + M (2 m + M) p_F^2] \notag\\&+
   2 m^3 M P p_F^4 V''_{2p_F}.
 \end{align}
%\end{widetext}
The scattering element (\ref{t1-1spec}) is proportional to temperature, contrary to temperature independent Eq.~(\ref{t11General}). Since the matrix element enters the expression for the force (\ref{force}) as $\left|t^{1-1}\right|^2$, the force gains two additional powers of temperature with respect to the previous result (\ref{TfourthGeneral}) and becomes proportional to $T^6$.

In order to evaluate the friction force, we were able to perform the  summations over momenta in Eq.~(\ref{force}) analytically. However we obtained a cumbersome expression, and for simplicity we state its form only in the limit of a very slow impurity $V\ll v_F$
\begin{align}\label{special1-1}
F^{1-1}=&
   -\frac{64\pi ^3 G^2  T^6 V  (V''_{2p_F})^2}{105  v_F^{10} \hbar ^5}.
\end{align}
The same conditions on temperature apply here as for Eq.~(\ref{TfourthGeneral}).

Comparing results (\ref{special1-1}) and (\ref{TfourthGeneral}), we find that the force (\ref{special1-1}) dominates over the force (\ref{TfourthGeneral}) in the parameter region
\begin{align}\label{condition5}
&\left|1 +\frac{m M \left(V_0-V_{2 p_F}+2 p_F V'_{2p_F}\right)}{G(M^2-m^2)} \right|<  4\pi\sqrt{\frac{2}{7}}mT\notag\\ &\times \frac{|V''_{2p_F}|}{\left|V_0-V_{2p_F}+2 p_F V'_{2 p_F}\right|}
\end{align}
Here for simplicity we considered a very slow impurity $V\ll v_F$.
We showed that the friction force can dramatically change its temperature dependence from the $T^4$ to $T^6$ law by tuning the system parameters. We will study the conditions (\ref{condition4}) and (\ref{condition5}) for some specific forms of the interaction potential in the following sections.

\subsection{$0-2$ processes}

Next we consider processes of the type $0-2$ illustrated in Fig.~1b, i.e.~we assume that all incoming and outgoing fermion momenta are positive. We introduce $p_r=p_1-p_2$ and $p_c=(p_1+p_2)/2$. In the same manner we define $q_r$ and $q_c$. We express $\delta P=Q-P=2(p_c-q_c)$ from the momentum conservation.  Analyzing  the product of Fermi distribution functions, we conclude that typical momenta are of the order of $|p_c-p_F|\sim |q_c-q_F|\sim T/v_F$, $|q_r|\sim |p_r|\sim T/v_F$. From the energy conservation we obtain that 
\begin{align}\label{p_c}
p_c=q_c+\frac{(q_r^2-p_r^2)}{8m( v_F-V)},
\end{align}
up to order $T^2$. Thus, contrarily to the $1-1$ processes, the typical momentum change $\delta P$ in the $0-2$ process scales quadratically with $T/E_F$. We further simplify the scattering matrix element (\ref{t}) as  
\begin{align}\label{eq:t02general}
t^{0-2}=Gp_r q_r \frac{G (M^2-m^2)-2 m M^3  V''_0(v_F-V)^2}{4L^2 m^2 M^3(v_F-V)^4}.
\end{align}
to the leading order in $T$. Note that the scattering matrix element changes the sign under transformations $p_r\to - p_r$ as well as  under $q_r\to -q_r$ as has to be the case due to anticommutation relations between the single particle fermion operators appearing  in the definition of $t$. 

Making use of Eq.~(\ref{p_c}) we express $\delta P$ and $p_c$ as functions of $q_c, q_r$ and $p_r$. The force is given by Eq.~(\ref{force}) and one needs to perform the summation over the remaining three momenta.  The phase space volume scales as $(T/v_F)^3$ since each summation is restricted to a range  of the width $T/v_F$, as discussed above. Thus, naively one would expect that  the force is proportional to $T^{9}$ since the typical impurity momentum change is $\delta P\propto T^2$ and the scattering matrix element $t^{0-2}\propto T^2$. However, the coefficient in front of $T^9$ vanishes. The reason is that the integrand behaves as an odd function under the exchange of $p_r$ and $q_r$ at this order in temperature.  The next order term in temperature in the expansion of the product of distribution functions appearing in Eq.~(\ref{force})  is nonzero and determines the force that scales as $T^{10}$. 
Indeed, we rewrite the energy conservation in the form
\begin{align}
\delta{(E_i-E_f)}\approx& \frac{1}{2 (v_F-V)}\delta\Big(p_c-q_c-\frac{1}{8m}\frac{q_r^2-p_r^2}{v_F-V}\Big),
\end{align}
by making use of the momentum conservation and then  carefully evaluate Eq.~(\ref{force}) to be
\begin{align}
F^{0-2}=&-\frac{4 \pi^7}{385 \hbar^5}\frac{G^2 V T^{10}}{m^6 M^6 v_F^{11}(v_F-V)^{11}}\notag\\ &\times\left[G(M^2-m^2)-2 m M^3  V''_0(v_F-V)^2\right]^2.
\end{align}
We assumed that temperature is low enough such that the following relation holds $T\ll E_F \text{min}\{1,v_F/V-1,\left(\frac{M}{m}\right)^{1/3}(1-V/v_F)^{2/3}\}$. We present some useful integrals in App.~\ref{appendix} that are needed for the evaluation of the force.

Next we consider $2-0$ processes. They are depicted in Fig.~1c.
Introducing new variables $p_i'=-p_i$,  $q_i'=-q_i$ for $i=1,2$ and $Q'=-Q$ we get
$F^{2-0}(P)=-F^{0-2}(-P)$. Thus 
the total force originating from the $0-2$ and $2-0$ processes, $F_T^{0-2}=F^{0-2}+F^{2-0}$, is
\begin{align}\label{FT02}
F^{0-2}_T=&-\frac{4 \pi^7}{385 \hbar^5}\frac{G^2 V T^{10}}{m^6 M^6 v_F^{11}}\notag\\ &\times\Bigg\{\frac{\left[G(M^2-m^2)-2 m M^3  V''_0(v_F-V)^2\right]^2}{(v_F-V)^{11}}\notag\\& +\frac{\left[G(M^2-m^2)-2 m M^3  V''_0(v_F+V)^2\right]^2}{(v_F+V)^{11}}\Bigg\}.
\end{align}

\subsection{Friction force}

Having analyzed different scattering processes and their contributions, we are now able to summarize the results for the friction force exerted on the impurity. The total force is the sum of two contributions
\begin{align}\label{eq:forceTotalGeneral}
F=F^{1-1}+F_T^{0-2}.
\end{align}
At low temperatures the leading contribution to the force is determined by the $1-1$ processes.  It is given by Eq.~(\ref{TfourthGeneral}) and scales\cite{castro_neto1996dynamics,gangardt2009bloch,matveev2012scattering} as $T^4$. However provided the condition (\ref{condition5}) is fulfilled the force scales as $T^6$ and is given by Eq.~(\ref{special1-1}). The processes $0-2$ and $2-0$ give rise to the friction force that scales as $T^{10}$ and may become the dominant ones by increasing the temperature.  In the following two sections, we apply these results to: i) the Tonks-Girardeau gas and  ii) strongly interacting bosons described by the Lieb-Liniger model.
In Sec.~\ref{screened} we examine the screened Coulomb interaction potential that does not satisfy the assumptions of this section.%%%%%%%%%%%%%%%%%%%%%%%%%%%%%%%%%%%
\section{Tonks-Girardeau gas}\label{noninteracting}
%%%%%%%%%%%%%%%%%%%%%%%%%%%%%%%%%%%

In this section we consider a one-dimensional system of hard-core bosons, known as a Tonks-Girardeau gas \cite{PhysRev.50.955,Girardeau}. The Hamiltonian $H_{TG}$ is a special case of the Lieb-Liniger model \cite{PhysRev.130.1605,PhysRev.130.1616}
\begin{align}\label{LiebLiniger}
H_{LL}=-\frac{\hbar^2}{2m}\sum_{i=1}^{N}\frac{\partial^2}{\partial x_i^2}+g\sum_{i<j}\delta(x_i-x_j).
\end{align}
By $m$ we denote the mass of bosons, while $g$ is the strength of the short-range repulsion.
$H_{TG}$ is realized for  $\gamma=mg/n\hbar^2 \to \infty$, where $n$ is the mean density of bosons. In addition, we consider an additional distinguishable particle, i.e.~an impurity, that interacts locally with bosons 
\begin{align}\label{impurity}
H_{i}=-\frac{\hbar^2}{2M}\frac{\partial^2}{\partial X^2}+G \sum_{j}\delta(X-x_j).
\end{align}
The total hamiltonian is given by
%\begin{align}\label{bosons}
$H_b=H_{TG}+H_{i}$.
%\end{align}
It is well known that the Tonks-Girardeau gas can be mapped\cite{Girardeau} onto spinless noninteracting fermions $H_{ff}=-\frac{\hbar^2}{2m}\sum_{i=1}^{N}\frac{\partial^2}{\partial x_i^2}$ . This mapping can be  extended to the case of Tonks-Girardeau gas with an impurity, as we show in App.~\ref{sec:mapping}. Thus, $H_b$ maps onto  $H_i+H_{ff}$, i.e.~onto Hamiltonian (\ref{Ham}) with $V_q=0$ and $G_q=G$.
For each eigenstate of the fermionic system there is a corresponding eigenstate in the bosonic system with the same energy.   Their wave functions are identical for one given arrangement of particles, e.g.~for $x_1<x_2<\ldots x_N$, while  for other arrangements their values are obtained using the (anti)symmetry of the wave function with respect to particle permutations\cite{Girardeau} and thus may differ up to a sign.
Namely,  the wave functions for fermions are antisymmetric, while those of bosons are symmetric.  As a consequence, a matrix element  between any two eigenstates of $H_b$ of any operator that does not permute bosonic particles,  coincide with its matrix element between the corresponding eigenstates of $H_i+H_{ff}$. Since the friction force can be written as $F= \langle \frac{d {P}}{dt} \rangle$, where $ P=-i\hbar \frac{\partial}{\partial X}$ is the impurity momentum  which  does not permute bosonic particles, the friction force exerted by noninteracting fermions coincides with  the friction force due to the Tonks-Girardeau bosons. In the following, we consider noninteracting fermions. Note that in this mapping, the sound velocity of bosons equals the Fermi velocity of fermions.

The results derived in the previous section apply to the system of noninteracting fermions and give the scattering matrix element originating from the $1-1$ process to be
\begin{align}
t^{1-1}=%&\frac{G^2 M (M^2-m^2)p_F^2}{L^2 (p_F^2-P^2)(M^2p_F^2-m^2P^2)}\\
& \frac{G^2 (M^2-m^2)}{L^2 M m^2(v_F^2-V^2)},
\end{align}
as follows from Eq.~(\ref{t11General}). The corresponding friction force (\ref{TfourthGeneral}) simplifies and reads as\cite{matveev2012scattering}
 \begin{align}\label{Tfourth}
F^{1-1}=&-\frac{2\pi}{15 \hbar^5}\frac{G^4 T^4 V (M^2-m^2)^2 (v_F^2+V^2)}{M^2 m^4 v_F^2(v_F^2-V^2)^5}.
\end{align}
For the $0-2$ processes the scattering matrix element (\ref{eq:t02general}) becomes 
\begin{align}
t^{0-2}&=\frac{G^2 (M^2-m^2)p_r q_r}{4L^2 m^2 M^3(v_F-V)^4},
\end{align}
and the total force originating from the $0-2$ and $2-0$ processes (\ref{FT02}) is
\begin{align}\label{T^10}
F_{T}^{0-2}=&-\frac{4\pi^7}{385\hbar^5} \frac{G^4T^{10} V  (M^2-m^2)^2}{m^6 M^6 v_F^{11}}\times \notag\\ &\left[\frac{1}{(v_F-V)^{11}}+\frac{1}{(v_F+V)^{11}}\right].
\end{align}
Note that condition (\ref{condition4}) simplifies to $M=m$. In this case the system is integrable  \cite{mcguire1965interacting,castella1993exact} and the scattering matrix element (\ref{t}) and the friction force (\ref{force}) vanish \cite{Zotos2002}.

The total friction force is  $F=F^{1-1}+F_T^{0-2}$. At low temperatures, the friction force is proportional to the fourth power of temperature\cite{castro_neto1996dynamics}. It is dominated by the $1-1$ type processes and is  given by Eq.~(\ref{Tfourth}). 
At higher temperatures $T>T^*$ the processes $0-2$ and $2-0$ may become more relevant. At crossover temperature  $T^*$, the expression (\ref{Tfourth}) is equal to the expression  (\ref{T^10}). However, the crossover temperature has to be sufficiently small and satisfy conditions given in the section \ref{sec:General}. For example, for very slow impurity, $V\ll v_F$, one gets the crossover temperature to be 
\begin{align}\label{crossoverTN}
T^*=\frac{2}{\pi}\left(\frac{77}{12}\right)^{1/6} E_F\left(\frac{M}{m}\right)^{2/3}.
\end{align}
Here the numerical prefactor $\frac{2}{\pi}\left(\frac{77}{12}\right)^{1/6}\approx 0.87$. For very light impurity ($M\ll m$) the crossover temperature satisfies the condition $T^*\ll E_F(M/m)^{1/3}$ required by (\ref{T^10}). However, the expression (\ref{Tfourth}) for $F^{1-1}$ is not valid for temperatures above $E_F M/m$ and its evaluation is beyond the scope of this paper. Thus we can not describe the crossover. To conclude, by decreasing the mass of the impurity and by increasing the temperature, the scattering processes $0-2$ gain more importance. However, for very slow impurity with mass of the order of $m$ or bigger, the friction force is given by $F_{1-1}$ at temperatures below the Fermi energy. In this case the processes $0-2$  can be neglected, since  $T^*$ becomes of the order of $E_F$ or bigger.

%%%%%%%%%%%%%%%%%%%%%%%
\section{Strongly interacting bosons}\label{strong}
%%%%%%%%%%%%%%%%%%%%%%%

In this section, we consider an impurity immersed in a one-dimensional system of bosons described by the Lieb-Liniger model \cite{PhysRev.130.1605,PhysRev.130.1616} (\ref{LiebLiniger}).
The construction of fermionic wave functions proposed in Ref.\cite{Girardeau} (see App.~\ref{sec:mapping} ) that leads to the mapping of Tonks-Girardeu bosons onto free spinless fermions can be used for an arbitrary interaction strength of bosons. It provides the exact mapping of the Lieb-Liniger  $H_{LL}$ model onto the Cheon-Shigehara model of fermions \cite{PhysRevLett.82.2536,Yukalov_2005}. The Hamiltonian of fermions can be written in the form\cite{PhysRevLett.99.110405} 
\begin{align}\label{Hfermions}
H_f=-\frac{\hbar^2}{2m}\sum_{i=1}^{N}\frac{\partial^2}{\partial x_i^2}-\frac{2\hbar^4}{m^2 g}\sum_{i>j}\delta''(x_i-x_j).
\end{align}
In Appendix~\ref{sec:mapping}, we show how this mapping can be extended in the presence of the impurity described by $H_{i}$ (\ref{impurity}). Thus,  $H_{LL}+H_{i}$ maps onto $H_f+H_{i}$ that in the second quantized representation is given by Eq.~(\ref{Ham}) with $V_q=2\hbar^2 q^2/m^2 g$  and $G_q=G$. We are interested in strongly interacting bosons with $\gamma=mg/n\hbar^2 \gg 1$ that corresponds to weakly interacting fermions. In the case of fermions $n$ denotes also the mean density. For convenience, we will consider the fermionic system in the following. However, the results for the friction force obtained in this section apply also to the system of Lieb-Liniger bosons with $\gamma\gg 1$.

 %%%%%%%%%%%%%%
\subsection{$1-1$ processes}
%%%%%%%%%%%%%%%%

We extend the analysis of the Sec.~\ref{noninteracting} to account for the interaction between the fermions, employing the results of Sec.~\ref{sec:General}. For the $1-1$ type processes, the matrix element is
\begin{align}\label{t1-1Strong}
t^{1-1}&=\frac{G^2 (M^2-m^2)}{L^2 M m^2(v_F^2-V^2)}\left( 1+\frac{8\pi}{\gamma} \frac{M m}{M^2-m^2} \frac{ \hbar v_F}{G}\right)\end{align}
in the leading order in temperature, as follows from Eq.~(\ref{t11General}). The corresponding force (\ref{TfourthGeneral}) becomes  
\begin{align}\label{fourthII}
F^{1-1}=&-\frac{2\pi}{15 \hbar^5}\frac{G^4 T^4 V (M^2-m^2)^2 (v_F^2+V^2)}{M^2 m^4 v_F^2(v_F^2-V^2)^5} \notag\\ &\times \left( 1+\frac{8\pi}{\gamma} \frac{M m}{M^2-m^2} \frac{\hbar v_F}{G}\right)^2.
\end{align}
We note that two contributions in the scattering element may have opposite sign. Thus one can tune the interactions such that the  scattering matrix element (\ref{t1-1Strong}) becomes zero. This happens for $g=g_c$ where
\begin{align}
g_c=8 \frac{\hbar^2 v_F^2  }{G }\frac{M m }{m^2-M^2},
\end{align}
provided $M\neq m$. We assume that background bosons repel each other, i.e., $g>0$. Thus, for an impurity lighter than the bosons ($M<m$) and that repels the bosons ($G>0$), or an impurity that is heavier than the bosons ($M>m$) and that attracts the bosons ($G<0$), this can be realized. However, this system is not integrable and the scattering matrix element is expected to be nonzero. Thus we evaluate it to the next order in $T/E_F$.  Using Eq.~(\ref{t1-1spec}), we obtain that it is linear in temperature and reads as
\begin{widetext}
\begin{align}
t^{1-1}=\frac{G^2 \left(m^2-M^2\right)}{2 L^2 m{p_F}^2 \left(M^2
   {p_F}^2-m^2 P^2\right)} \left[-\delta P
   P (m+M)^2+2 M ({p_1}-{p_F}) (m {p_F}+M
   P)+2 M ({p_2}+{p_F}) (M P-m
   {p_F})\right],
\end{align}
for $g=g_c$. Contrary to the results of Sec.~\ref{sec:General} where the friction force is given only for a very slow impurity, here we state its general form
\begin{align}\label{forceNew}
F^{1-1}=&-\frac{4 \pi ^3 G^4 T^6 V (m^2-M^2)^2}{315 \hbar ^5m^{10} M^2
   v_F^{14}  (v_F^2-V^2)^5}\times \notag\\&\left\{{5 \left(v_F^2-V^2\right)^4
   f\left({v_F}/{V}\right) \left[m^4
   \left(v_F^2+V^2\right)+M^4 V^2\right]}+7{v_F^6}
   \left(v_F^2+V^2\right) \left(m^4 v_F^2+M^4 V^2\right)\right\}.
\end{align}
\end{widetext}
Here, the function $f$ is defined as
\begin{align}
\label{eq:f2}
f(a)=&\frac{21 a^7}{64 \pi ^6}\int_{-\infty}^{\infty}\frac{e^{-x}x^5dx}{\cosh(x)-\cosh{(a x)}},\quad\quad a>1.
\end{align}
It has the following limiting behavior $\lim_{a\to \infty}f(a)= 1$. 
Comparing the expressions (\ref{fourthII}) and (\ref{forceNew}), we find that expression (\ref{forceNew}) is the leading contribution to the force provided
\begin{align}\label{eq:condition}
\left|1-\frac{g_c}{g}\right|<2 \sqrt{\frac{2}{7}} \pi\frac{T}{m v_F^2}.
\end{align}
Here, for simplicity, we considered the case of very slow impurity $V\ll v_F$.

%%%%%%%%%%%%%%%%%%%
\subsection{$0-2$ processes}
%%%%%%%%%%%%%%%%%%%

Next we consider the $0-2$ processes.
From Eq.~(\ref{eq:t02general}) follows that the scattering matrix element reads as 
\begin{align}\label{tsi}
t^{0-2}=&\frac{G^2 (M^2-m^2) p_r q_r}{4L^2 m^2 M^3(v_F-V)^4}\left[1-\frac{8\pi \hbar M^3 (v_F-V)^2}{\gamma G m v_F (M^2-m^2)}\right]
\end{align}
in the lowest order in temperature. Note that by tuning the interactions such that Eq.~(\ref{t1-1Strong}) vanishes, the matrix element (\ref{tsi}) does not vanish but its two contributions rather sum up leading to an increase of the friction force with respect to the noninteracting case.
%The force is
%\begin{align}
%F^{0-2}
%=&-\frac{16\pi^7}{385\hbar^5} \frac{T^{10} V G^4 }{v_F^{11}(v_F-V)^{11}}\frac{(M^2-m^2)^2}{m^6 M^6}\times \notag\\ &\left[1-\frac{8\pi \hbar M^3 (v_F-V)^2}{\gamma G m v_F (M^2-m^2)}\right]^2
%\end{align}
The force $F_T^{0-2}=F^{0-2}+F^{2-0}$ is given by
\begin{align}
\label{T^10II}
F_T^{0-2}=&-\frac{4\pi^7}{385\hbar^5} \frac{G^4 T^{10} V (M^2-m^2)^2}{m^6 M^6v_F^{11}}\times \notag\\ &\Bigg\{ \frac{1}{(v_F-V)^{11}}\left[1-\frac{8\pi \hbar M^3 (v_F-V)^2}{\gamma G m v_F (M^2-m^2)}\right]^2\notag\\&+\frac{1}{(v_F+V)^{11}}\left[1-\frac{8\pi \hbar M^3 (v_F+V)^2}{\gamma G m v_F (M^2-m^2)}\right]^2\Bigg\}.
\end{align}

%%%%%%%%%%%%%%%%%%%
\subsection{Friction force}
%%%%%%%%%%%%%%%%%%%

The friction force is given by Eq.~(\ref{eq:forceTotalGeneral}). At low temperatures, it is dominated by the $1-1$ processes. In a general case, it is proportional to the fourth power of temperature $T^4$ and is  given by Eq.~(\ref{fourthII}). However, if the condition (\ref{eq:condition}) is satisfied, the force is given by Eq.~(\ref{forceNew}) and scales as the sixth power of the temperature. At higher temperatures $T>T^*$ the processes $0-2$ and $2-0$ may become dominant. Here the crossover temperature  $T^*$ is defined such that two contributions in Eq.~(\ref{eq:forceTotalGeneral}) are equal. In order that the crossover takes place, the crossover temperature has to be within the validity of the theory and satisfy the constraints discussed in Sec.~\ref{sec:General}.

For example, for a very slow impurity $v_F\gg V$ assuming that the condition (\ref{eq:condition}) is not satisfied, one gets 
\begin{align}\label{crossoverT}
T^*= \frac{2}{\pi}\left(\frac{77}{12}\right)^{1/6}  E_F \left(\frac{M}{m}\right)^{2/3}\left|\frac{1-g_c/g}{1+g_c M^2/g m^2}\right|^{1/3}.
\end{align}
Here the numerical prefactor $\frac{2}{\pi}\left(\frac{77}{12}\right)^{1/6}\approx 0.87$.
As for noninteracting fermions, the processes $0-2$ are more important for light impurities provided the interactions between the fermions do not dominate over the impurity--liquid coupling, i.e.~$|g_c| M^2\lesssim |g| m^2$.  We can tune the interactions such that $g$ is in the vicinity of $g_c$, in order to further decrease the force $F^{1-1}$.
However, Eq.~(\ref{fourthII}) ceases to be valid at the  crossover temperature. For example, for light impurity $M\ll m$ and the interactions such that $g_cM^2\ll g m^2$, the crossover temperature becomes  $T^*/E_F\approx \frac{M^{2/3}}{m^{2/3}}\left|{1-g_c/g}\right|^{1/3}\ll 1$.  The condition (\ref{eq:condition}) implies $\left|{1-g_c/g}\right|>M/m$ and thus $T^*/E_F>M/m$ that is out of the validity of the expression (\ref{fourthII}). Further, note that for $M>m$ and repulsive interactions $g>0$ and $G>0$, the crossover temperature (\ref{crossoverT}) is greater than the Fermi energy, and thus the friction force is
\begin{align}\label{force1}
F=&-\frac{2\pi}{15 \hbar^3}V\frac{G^2T^4}{m^2v_F^{8}}\left( \frac{G}{\hbar v_F}\frac{M^2-m^2}{m M}+8\frac{\hbar v_F}{g}\right)^2,
\end{align}
for all temperatures $T\ll E_F$. 

If the interaction between fermions dominates over the impurity coupling such that $|g_c/g|\text{min}\{1, M^2/m^2\}\gg 1$ we can neglect the $F^{0-2}_T$ contribution for all temperatures below the Fermi energy.

If the interactions are tuned such that Eq.~(\ref{eq:condition}) holds, we should compare Eq.~(\ref{forceNew}) and  Eq.~(\ref{T^10II}).
We get the crossover temperature
\begin{align}
T^*=\frac{2}{\pi}\left(\frac{22}{3}\right)^{1/4} E_F \frac{{M}/{m}}{\sqrt{1+M^2/m^2}},
\end{align}
that is below the Fermi energy for very light impurities. 
However, Eq.~(\ref{forceNew}) breaks down in this region while Eq.~(\ref{T^10II}) remains valid.  To conclude, at temperatures below the Fermi energy,  the processes $0-2$ may become the dominant ones only  for  very light impurities at temperatures  $T\gtrsim E_F M/m$ provided $g$ is of the order of $| g_c|$.

%%%%%%%%%%%%%%%%%%%%%%%
\section{Screened Coulomb interaction}\label{screened}
%%%%%%%%%%%%%%%%%%%%%%%

In this section we  consider fermions interacting via the screened Coulomb interaction. This case  is relevant for quantum wires. The potential is given by $V(r)=e^2/|r|-e^2/\sqrt{r^2+ 4d^2}$,  where $d$ denotes the distance between a wire and a conducting plane representing the gate, while $e$ denotes the electron charge. The Fourier transform of the potential has the form
\begin{align}\label{SC}
V_q=2 e^2 \ln{\left(\frac{d}{w}\right)}-2 e^2\frac{q^2 d^2}{\hbar^2}\ln{\left(\frac{\hbar}{|q|d}\right)}
\end{align}
for low momenta $|q|\ll \hbar/d \ll \hbar/w$ and within the logarithmic accuracy\cite{ZoranKostya}. Here $w$ denotes the width of the wire.

The first term in (\ref{SC})  describes the contact interaction. It does not contribute to the matrix element since the contact interaction that does not have any effect on the wave function of fermions that is antisymmetric with respect to permutations of particle's coordinates. We now check that this property holds in our theory. The contribution in $t$ originating from the two terms linear in $V$ in Eq.~(\ref{t}) should vanish for $V_q=V$. The first term gives vanishing contribution after action of  the operator $1-\hat{\mathcal{A}}(p_1,p_2)$, since it  is symmetric under exchange of $p_1$ and $p_2$. Similarly, the second one (i.e. the second term of Eq.~(\ref{t})) vanishes after action of the operator $1-\hat{\mathcal{A}}(q_1,q_2)$. So, there is no  contribution coming from the contact interaction between the fermions in $t$ as has to be the case. Thus only the second term in Eq.~(\ref{SC}) causes the scattering. Note that we consider weak interactions $G\ll \hbar v_F$ and $e^2\delta^2|\ln(\delta)|\ll \hbar v_F$, where $\delta=dp_F/\hbar\ll 1$.

We point out that the Fourier transform of the interaction potential  (\ref{SC}) is not an analytic function of momentum and thus the theory presented in Sec.~{\ref{sec:General}} does not apply here. Nevertheless, the dependence of the fermion and the impurity momenta on temperature remains the same  as in Sec.~{\ref{sec:General}} as well as the temperature constraints. In this section  we evaluate the scattering matrix element (\ref{t}) and the friction force (\ref{force}) for the screened Coulomb interaction potential. 

\subsection{$1-1$ processes}

We start by considering the 1-1 processes. We determine the matrix element (\ref{t}) to be
\begin{align}\label{tSC}
t^{1-1}=&\frac{G^2(M^2-m^2)}{L^2 M m^2 (v_F^2-V^2)}\left[1+8\frac{\delta^2 e^2}{G}\frac{M m}{M^2-m^2}\ln{(2E\delta)}\right],
\end{align} 
in the leading order.
Here $E$ denotes the Euler's number. 
The first term in Eq.~(\ref{tSC}) describes noninteracting fermions, while the second term describes the effect of the interaction. We calculate the friction force
\begin{align}\label{TfourthSC}
F^{1-1}=&-\frac{2\pi}{15 \hbar^5}\frac{G^4 T^4 V (M^2-m^2)^2 (v_F^2+V^2)}{M^2 m^4 v_F^2(v_F^2-V^2)^5}\notag\\&\times \left[1+8\frac{\delta^2 e^2}{G}\frac{M m}{M^2-m^2}\ln{(2E\delta)}\right]^2.
\end{align}
By analyzing the expression (\ref{TfourthSC}), wee see that its two contributions can cancel each other. This happens for $G=G_c$ where 
\begin{align}
G_c
%\frac{8 \delta ^2 e^2 m M \ln (2 E \delta )}{m^2-M^2}\\
=8 \delta ^2 e^2\ln (2 E \delta )\frac{ m M }{m^2-M^2},
\end{align}
provided $M\neq m$. The cancellation can be realized for a repulsive interaction between the impurity  and the liquid ($G>0$) if the impurity is heavier than the background particles $M>m$, and in the opposite case $M<m$ for  an attractive interaction $G<0$. This  should be contrasted with the case of strongly interacting repulsive bosons considered in the previous section where the related cancellation of the scattering matrix element 1-1 takes place for $G (m-M)>0$.

Next we evaluate the scattering matrix element (\ref{t})  for $G=G_c$  and obtain that it is proportional to temperature
%\begin{widetext}
\begin{align}
t^{1-1}=&-\frac{16 \delta ^4 e^4 m M^2 \ln (2 E \delta )}{L^2
  p_F^4 \left(m^2-M^2\right) (M^2 p_F^2-m ^2P^2) }\notag\\ &\times\left[\alpha (p1-p_F)+\beta (p2+p_F)+\gamma \delta P\right],\\
\alpha=& \left[-p_F \ln{\left(4 \delta ^2\right)}
   (m p_F+M P)-3 m p_F^2-2 M P p_F\right]\notag\\ &\times2 M p_F,\\
\beta=& \left[p_F \ln{\left(4 \delta ^2\right)}
   (m p_F-M P)+3 m p_F^2-2 M P p_F\right]\notag\\ &\times2 M p_F,\\
\gamma=&P p_F^2\left[ \ln{\left(4 \delta
   ^2\right)} (m+M)^2+4 m M +3 m^2 +2 M^2
   \right].
\end{align}
%\end{widetext}
For simplicity, we state the friction force for a very slow impurity $V\ll v_F$ only
\begin{align}\label{F11SCS}
F^{1-1}= -\frac{16384 \pi ^3 \delta ^8 e^8 m^{12} M
   P T^6 \ln ^2(2 E \delta ) \ln^2{\left(4 E^3 \delta
   ^2\right)}}{105 p_F^{14} \hbar ^5\left(m^2-M^2\right)^2}.
\end{align}
Note that for special case $M=m$ this result does not apply since Eq.~(\ref{TfourthSC}) cannot vanish and is always the leading contribution to $F^{1-1}$. Further, we point out that Eq.~(\ref{F11SCS}) is the leading contribution for the parameter range
\begin{align}\label{conditionSC}
\left| 1-G_c/G\right|< \sqrt{\frac{1}{14}} \pi\frac{T}{E_F}\frac{\ln{\left(4 E^3 \delta
   ^2\right)}}{\ln{(2 E \delta )}}.
\end{align}
Note that Eq.~(\ref{SC}) is determined within the logarithmic accuracy, and thus one can neglect all the numerical factors under logarithms in this section. However, we kept them for completeness of the solution for the given model potential (\ref{SC}). 

\subsection{$0-2$ processes} 

Next we consider the $0-2$ processes. They lead to the matrix element
%\begin{widetext}
\begin{align}\label{t0-2SC}
t^{0-2}
=&\frac{G^2 (M^2-m^2)p_r q_r}{4L^2m^2M^3 (v_F-V)^4}+\frac{d^2e^2 G }{2 L^2 \hbar^2m (v_F-V)^2}\notag\\ &\times
\Big[(p_r-q_r)^2\ln{(|p_r-q_r|d/\hbar)}\notag\\ &-(p_r+q_r)^2\ln{(|p_r+q_r|d/\hbar)}+4 p_r q_r(\ln{2}-1)\Big].
\end{align}
%\end{widetext}
The total force originating form the $0-2$ and $2-0$ processes reads as 
\begin{align}\label{T^10SC}
F_{T}^{0-2}=&-\frac{4\pi^7}{385\hbar^5} \frac{G^4T^{10} V  (M^2-m^2)^2}{m^6 M^6 v_F^{11}}\times \notag\\ &\Bigg\{\frac{\left[1-\frac{8 \delta^2 e^2 M^3(v_F-V)^2\ln{\left(\frac{T d}{\hbar v_F}\right)}}{G m v_F^2 (M^2-m^2)}\right]^2}{(v_F-V)^{11}}
\notag\\&+\frac{\left[1-\frac{8 \delta^2 e^2 M^3(v_F+V)^2\ln{\left(\frac{T d}{\hbar v_F}\right)}}{G m v_F^2 (M^2-m^2)}\right]^2}{(v_F+V)^{11}}\Bigg\}
\end{align}
within the logarithmic accuracy. The functional dependence on momenta of the contribution originating from the screened Coulomb interaction in (\ref{t0-2SC})  is different from previously considered cases. If it dominates over the contribution coming from the impurity-liquid coupling, the force $F_{T}^{0-2}$ becomes proportional to $T^{10}\ln^2{\left(\frac{T d}{\hbar v_F}\right)}$. 
Also, if one tunes $G$ to be equal to $G_c$, the friction force $F^{0-2}_T$ becomes bigger than the corresponding one in the case of noninteracting fermions (\ref{T^10}).

\subsection{Friction force}

As expected, at low temperatures the $1-1$ type processes are the dominant ones and the friction force is given by Eq.~(\ref{TfourthSC}). If however the condition (\ref{conditionSC}) is fulfilled, the force scales as sixth power of temperature and  is given by (\ref{F11SCS}). Increasing the temperature the processes $0-2$ and $2-0$ gain in importance. We define the crossover temperature  $T^*$ such that the force $F^{1-1}$ equals $F^{0-2}_T$. However, the crossover temperature is out of the domain of validity of our theory, as we discuss below. 

For example, assuming that the condition (\ref{conditionSC}) is not satified, in the case of very slow impurity $v_F\gg V$, one gets 
\begin{align}\label{crossoverTSC}
T^*=\frac{2}{\pi}\left(\frac{77}{12}\right)^{1/6} E_F\left(\frac{M}{m}\right)^{2/3}\left|\frac{1-G_c/G}{1+\mathcal{N}G_c M^2/G m^2}\right|^{1/3},
\end{align}
where $\mathcal{N}\approx1+\ln{(T^*/E_F)}/\ln{\delta}$ and $\frac{2}{\pi}\left(\frac{77}{12}\right)^{1/6}\approx 0.87$. If the interaction between the fermions dominates over  the impurity-liquid coupling such that $|G_c/G|\text{min}\{1,\mathcal{N}M^2/m^2\}\gg 1$, the $0-2$ processes are always negligible. This is the case also for a heavy impurity $M>m$ and an attractive impurity-liquid interaction $G<0$. The reason is that in these cases $T^*\sim E_F$ and thus the $0-2$ processes can be safely neglected in the considered parameter region. 

If the interaction between the fermions does not dominate over the impurity-liquid coupling, in the case of heavy impurity $M>m$ and repulsive interactions, in order that the crossover temperature  satisfies $T^*/E_F\ll 1$, we should tune the interactions such that $G$ is in close vicinity of $G_c$. Then, the condition (\ref{conditionSC}) becomes valid and, as we show below, the scatterings $0-2$ are negligible. Indeed, in a general case, by comparing Eq.~(\ref{F11SCS}) with Eq.~(\ref{T^10SC}) we find
\begin{align}
T^*=\frac{2}{\pi }\left(\frac{11}{6}\right)^{1/4}E_F \frac{M}{m}\left(\frac{|\ln{(4 E^3 \delta)}|}{|\ln{(2E\delta)}+ M^2\ln{(\delta c)}/m^2|}\right)^{1/2}
\end{align}
where $c$ is of the order of  ${T^*/E_F}$. To conclude, for heavy impurities $M>m$,  the $0-2$ processes may be neglected for temperatures  below the Fermi energy. Evenmore, the $0-2$ processes may be important only for very light impurities ($M\ll m$) provided $|G|$ of the order of $|G_c|$ and at temperatures above $E_F M/m$ , where an analysis of $F^{1-1}$ is missing.

%%%%%%%%%%%%%%%%%%%%%%%%
\section{Conclusions and discussion}\label{conclusions}
%%%%%%%%%%%%%%%%%%%%%%%%

In this paper we have studied the scattering of a slow quantum impurity
off background particles in one-dimensional liquids by employing the microscopic description. In particular, we have focused on the friction force exerted on the impurity. We have determined its dependence on the system parameters for several interacting systems: (i) assuming the Fourier transform of the interaction potential felt by  fermions is an analytic function of momenta and (ii) for fermions interacting via screened Coulomb interaction. Some special cases of the class (i) have been studied in detail: (a) the Tonks-Girardeau gas or equivalently  the system of  non-interacting fermions and (b) bosons interacting strongly via the contact interaction or the Cheon-Shigehara model for weakly interacting fermions.  We have considered temperatures well below the Fermi energy.

We have demonstrated that by tuning the system parameters the expected $T^4$ dependence of the friction force\cite{castro_neto1996dynamics,gangardt2009bloch,matveev2012scattering}  can dramatically change into a  new universal $T^6$ law as shown by Eqs.~(\ref{special1-1}, \ref{forceNew}, \ref{F11SCS}).  These findings allow us to design the desired friction force in the system. 
It would be interesting to verify if this result remains valid for  arbitrary strength of the interaction between the background particles. 
Note that the friction force on the impurity in the system of weakly interacting bosons has been studied in Ref.~\cite{PRLimpurity}. It was found that the force can become proportional to $T^8$ due to scattering off Bogoliubov quasiparticles. However, this result ceases to be valid at very low temperatures $T\ll mv^2\gamma^{1/4}$, where $\gamma\ll 1$.  Here $v$ denotes the sound velocity. The reason is that the low-energy quasiparticles are of fermionic nature \cite{RevModPhys.84.1253}.  
Note that at arbitrary interaction strength the region of low-energy fermionic quasiparticles becomes wider \cite{ZoranPRL} and thus their scattering off the impurity determines the low-temperature friction force. 
Also, the system of fermions at arbitrary interaction and at low temperatures can be described by  weakly interacting fermionic quasiparticles  by applying the refermionization procedure\cite{Rozhkov}.

While the above discussed friction force is the result of the scattering of the impurity off the fermions situated around different Fermi points, we have analyzed also the related process where the fermions are around the same Fermi point. They lead to the $T^{10}$ contribution of the friction force, and can become dominant at higher temperatures only for very light impurities.  

Further, in this paper we have studied a mobile impurity weakly coupled to the system, $G\ll \hbar v_F$. However, using the Luttinger liquid description, in Ref.~\cite{castro_neto1996dynamics} it was shown that a strongly coupled slow impurity moving in a system of repulsive fermions is well described by a weak-coupling impurity approach at low temperatures: $i)$  $T< E_F m/M$ for a  very heavy impurity $M\gg m$  and $ii)$ $T <E_F$ for very light impurity $M\ll m$. These findings imply that our results for the friction force from sections \ref{sec:General}, \ref{strong} and \ref{screened} apply also for the impurity strongly coupled to the background particles.  Note that the friction force is expected to be temperature independent for a very heavy $M\gg m$ and strongly coupled impurity for temperatures above $E_F m/M$\cite{castro_neto1996dynamics}.

We point out that our results indicate that there is universal dependence of the low temperature $T^4$ friction force on the impurity velocity in one-dimensional systems as shown by Eqs.~(\ref{TfourthGeneral},\ref{Tfourth},\ref{fourthII},\ref{TfourthSC}). This statement is in agreement with the findings of  Ref.~\cite{PRLimpurity} that studies bosons interacting weakly via contact interaction. In that case  the sound velocity of a system of bosons corresponds to the Fermi velocity.

Versatile and highly controlled experimental realizations with ultra cold gases provide promising platforms where our results could be tested. The impurities can be realized by mixing different atoms \cite{PhysRevLett.109.235301,catani2012quantum}, exciting a few atoms into a different hyperfine internal state \cite{chikkatur2000suppression,palzer2009quantum} or with trapped ions \cite{zipkes2010trapped,schmid2010dynamics}. Even more important for our work is the possibility to tune the interactions using the Feshbach resonance\cite{RevModPhys.82.1225} in order to observe the predicted dramatic change of the friction force.

\section*{Acknowledgements}
We thank Zoran Ristivojevic for helpful discussions and his interest in the initial stage of this work.

%%%%%%%%%%%%%%%%%%%%%%%
\appendix
\section{Useful integrals}\label{appendix}
%%%%%%%%%%%%%%%%%%%%%%%

Some useful integrals needed for the evaluation of the friction force are presented in this appendix. For analysis of the $1-1$ type processes, we calculated
%\begin{align}
%&\int_0^{\infty}  \frac{dx}{\cosh{(x-a)}\cosh{(x-b)}}=\notag \\ &=\frac{[a+\ln{(\cosh{a})}-\ln{(1-\tanh{b}})]}{\sinh{(a-b)}},
%\end{align}
\begin{align}
\int_{-\infty}^{+\infty}dx \frac{x^3 e^{-x}}{\cosh{(x)}-\cosh{(a x)}}=\frac{16 \pi^4}{15}\frac{a(a^2+1)}{(a^2-1)^4} , \quad a>1.
\end{align}
In the case of the $0-2$ processes, we evaluated the integrals 
\begin{align}
&\int_{0}^{\infty}dx \frac{\left[\mathrm{L}_5(-e^{-x})-\mathrm{L}_5(-e^{x})\right]^2}{\sinh^2{(x})}=\frac{97 \pi^{10}}{665280},\\
&\int_{0}^{\infty} dx\frac{x (\pi^2+x^2)\left[\mathrm{L}_7(-e^{-x})-\mathrm{L}_7(-e^{x})\right]}{\sinh^2{(x})}=\frac{193\pi^{10}}{277200}.
\end{align}
Here $\mathrm{L}_n(x)$ denotes the polylogarithm function.
%%%%%%%%%%%%%%%%%%
\section{Fermion-boson duality}\label{sec:mapping}
%%%%%%%%%%%%%%%%%%

The mapping\cite{Girardeau}  of the Tonks-Girardeau (TG) gas to noninteracting fermions can be performed in the following way. We denote by $\Psi_B(x_1,\ldots x_N)$ a many-body wave function of an eigenstate of the TG Hamiltonian of $N$ bosons with coordinates $x_1,\ldots x_N$. Then, the many-body wave function of the corresponding eigenstate of spinless noninteracting fermions is given by
\begin{align}\label{mapping}
\Psi_F(x_1,\ldots,x_N)=\prod_{i>j}^N\text{sgn}(x_i-x_j)\Psi_B(x_1,\ldots,x_N),
\end{align} 
where $\text{sgn}(x)=x/|x|$. It has the same energy as the corresponding eigenstate of bosons. The product of terms $\prod_{i>j}^N\text{sgn}(x_i-x_j)$ accounts for the antisymmetry of the wave function with respect to particle permutations. 

We point out that in the case of TG gas with an impurity, the mapping onto noninteracting fermions with an impurity can be easily generalised. The reason is that the impurity is distinguishable from the particles of the system. The mapping can be written as
\begin{align}\label{mappingImpurity}
\Psi_F(X,x_1,\ldots,x_N)=\prod_{i>j}^N\text{sgn}(x_i-x_j)\Psi_B(X,x_1,\ldots,x_N),
\end{align} 
where $X$ denotes the impurity coordinate while $\Psi_B(X,x_1,\ldots,x_N)$ is a many-body wave function of an eigenstate of the system of TG bosons and the impurity described by (\ref{impurity}). The contact interaction of the impurity with bosons imposes  continuity of $\Psi_B$ and discontinuity of its partial derivative $\partial_{x_i}$ at $x_i=X^{\pm}$. Here $X^{\pm}=\lim_{\varepsilon\to 0^+}X\pm \varepsilon$. This constraint remains unchanged and applies also to the corresponding $\Psi_F$ given by (\ref{mappingImpurity}). That is why the interaction potential of the impurity is $G\sum_i\delta (X-x_i)$ both in the case of bosons and fermions.

If the mapping (\ref{mapping}) is applied onto the Lieb-Liniger model at arbitrary interaction strength, one obtains a system of fermions in a specific interaction potential\cite{PhysRevLett.82.2536,Yukalov_2005}. Namely, the interaction potential $ \sum_{i<j}g\delta(x_i-x_j)$ imposes the continuity of the $\Psi_B$ wave function but the discontinuity of $(\partial_{x_i}-\partial_{x_j})\Psi_B$ at $x_i=x_j^{\pm}$. These constraints on $\Psi_B$ imply different constraints on the corresponding $\Psi_F$ function. By its construction the wave  function of fermions is antisymmetric under particle permutations and thus it is discontinuous while the combination of the partial derivatives $(\partial_{x_i}-\partial_{x_j})\Psi_F$  is a continuous function at $x_i=x_j^{\pm}$ for a finite nonzero value of $g$. These constraints can be incorporated into the Hamiltonian of the system  by a specific interaction potential\cite{PhysRevLett.82.2536,Yukalov_2005}, that can be written as\cite{PhysRevLett.99.110405}  $-2\delta''(x_i-x_j)/m^2 g$. The interaction strength is inversely proportional to the interaction strength of the bosons, and thus in the TG limit we get free fermions.
We further point out that the mapping (\ref{mappingImpurity}) also applies to the Lieb-Liniger model with the impurity (\ref{impurity}) giving the system of fermions modeled by Hamiltonian (\ref{Hfermions}) with the impurity described by Eq.~(\ref{impurity}).

\end{document}